\documentclass[12pt]{article}
\pdfoutput=1
\usepackage{hyperref}
\usepackage{cite}
\usepackage{color}
\usepackage{graphicx}
\usepackage{amsmath}
\usepackage{amssymb}
\usepackage{xspace}
\usepackage{soul}
\usepackage{tikz}

\makeatletter
\@addtoreset{equation}{section}

\makeatletter
\renewcommand\section{\@startsection {section}{1}{\z@}%
                                   {-3.5ex \@plus -1ex \@minus -.2ex}
                                   {2.3ex \@plus.2ex}%
                                   {\normalfont\large\bfseries}}
\renewcommand\subsection{\@startsection{subsection}{2}{\z@}%
                                     {-3.25ex\@plus -1ex \@minus -.2ex}%
                                     {1.5ex \@plus .2ex}%
                                     {\normalfont\bfseries}}

\newcommand{\be}{\begin{equation}}
\newcommand{\ee}{\end{equation}}
\newcommand{\beq}{\begin{eqnarray}}
\newcommand{\eeq}{\end{eqnarray}}

\newcommand{\gone}[1]{{}}

\newcommand{\A}{{\cal A}}

\begin{document}
\begin{titlepage}

\rule{0ex}{0ex}

\vfil

\begin{center}

{\bf \Large
Weighted average over Narain moduli space \\ as   $T \bar T$ deformation of  CFT target space
}

\vfil

Soumangsu Chakraborty$^1$,  Akikazu Hashimoto$^2$

\vfil

{}$^1$ Department of Theoretical Physics,\\
Tata Institute for Fundamental Research,\\
$1^{st}$ Homi Bhabha Road, Mumbai 400005, India

{}$^2$ Department of Physics, University of Wisconsin,\\ Madison, WI 53706, USA

\vfil

\end{center}

\begin{abstract}
\noindent We consider the weighted average of a  two dimensional  CFT, whose target space is $T^2$, over its Narain moduli space. We take as the weighing function the integral kernel which gives rise to $T \bar T$ deformation when applied to the world sheet moduli data of the partition function viewed as vacuum amplitude when the world sheet is a torus. We compute the smeared partition function where this kernel is applied to the target space moduli. Smearing the partition function over the parameter space of a field theory generally leads to the breakdown in the ability to write the partition function as a sum over Boltzmann factor with unit coefficients. The weight function inspired by the $T \bar T$ deformation appears to be an exception to this general expectation. We show that this smearing leads to a marginal deformation corresponding to the overall rescaling of the target space $T^2$. 
\end{abstract}
\vspace{0.5in}

\end{titlepage}

\section{Introduction}

Recently, there have been some interest in the subject of averaging field theories over the space of parameters. In \cite{Maloney:2020nni,Afkhami-Jeddi:2020ezh,Datta:2021ftn,Benjamin:2021wzr}, a setup where averaging $2d$ CFT's whose target space is $T^d$ over the Narain moduli space $O(d,d,Z)\backslash O(d,d)/(O(d)\times O(d))$ has been discussed.  In \cite{Maloney:2020nni,Afkhami-Jeddi:2020ezh,Datta:2021ftn,Benjamin:2021wzr}, the averaging being performed is uniform and the integration measure is invariant under $O(d,d,R)$.  It is straight forward to compute the average of observables such as the vacuum world sheet torus amplitude.\footnote{We use the term ``world sheet'' to refer to the two dimensional space on which the CFT is defined even though we are not always working in the context of string theory.} Even though the CFT vacuum torus amplitude is interpretable as a Boltzmann sum, this feature is lost in the averaged quantity. This is not unexpected since in field theories, smearing\footnote{We use the term ``smearing'' to refer to the process of integrating over the parameter space with some  weighing factor.}  generically breaks locality, causality, and unitarity. Smearing of field theories have also been considered in phenomenological contexts e.g.\ \cite{Coleman:1988tj,Balasubramanian:2020lux} and in a more mathematical context in \cite{Dymarsky:2020pzc}. The relation between smearing and the universality class of random matrix ensembles was discussed recently in \cite{Afkhami-Jeddi:2021qkf}.  Although the work of  \cite{Maloney:2020nni,Afkhami-Jeddi:2020ezh,Datta:2021ftn,Benjamin:2021wzr} was motivated by the interest in understanding the holographic relation between an ensemble average of a CFT on the boundary and the quantum gravity theory on the bulk, we will only consider the effects of smearing form the field theory point of view in this article.

In a seemingly unrelated topic, there has also been some interesting developments in the subject of $T \bar T$ deformed field theories \cite{Smirnov:2016lqw,Cavaglia:2016oda}. The basic construction consists of a QFT in two dimensions, deformed by an irrelevant operator $\mu (T_{zz} T_{\bar z \bar z} - T_{z \bar z}^2)$ where $\mu$ is the irrelevant deformation parameter. To see the salient features of this deformation, it suffices to start with a CFT. Usually, deformation by an irrelevant operator does not lead to a well defined quantum system without additional information needed to specify the UV fixed point. In the case where the deformation is precisely of the type considered by \cite{Smirnov:2016lqw,Cavaglia:2016oda}  living on a two dimensional space time of the form $R \times S$, it was shown that the energy spectrum can be determined unambiguously. The resulting spectrum was found to exhibit Hagedorn behavior, indicating that the deformed system is non-local. Partition function is a simple observable which encodes the spectrum of the theory and was studied in the context of $T \bar T$ deformation by several authors \cite{Cardy:2018sdv,Dubovsky:2018bmo,Aharony:2018bad,Hashimoto:2019wct}.  The partition function was found to satisfy a diffusion type flow equation with respect to $\mu$ and was solvable in terms of action of an integral kernel where the initial condition at $\mu=0$ is set by the partition function of the undeformed theory. In other words, partition function of $T \bar T$ deformed field theories can be obtained by starting with the undeformed partition function and applying a smearing procedure over the data specifying the moduli of the world sheet torus $(\Omega, \tau, \bar \tau)$ where $\tau$ is the complex structure and $\Omega$ parameterizes the size.\footnote{The dependence on $\Omega$ decouples if the theory on the torus is a CFT.}

In this article, we explore the following simple question: what happens if we start with a $2d$ CFT whose target space is $T^2$, and apply the $T\bar T$ smearing not to the world sheet but rather the target space $T^2$. The moduli of target space $T^2$ is parameterized by complex structure $\xi$ and K\"ahler structure $\rho$. The $T \bar T$ smearing can naturally act on $ \xi$, $\bar \xi$, and  $\rm{Im}\, \rho$. From the perspective of averaging over the Narain moduli space, we are prescribing a specific weight which is no longer uniform. There is some freedom left in prescribing the $\rm{Re}\,  \rho$ dependence in the weighing factor.\footnote{$\mbox{Re}\, \rho$ is the coefficient of the topological term in the action, corresponding to $B$-field in the context of string theory. } The main result we report in this article is that for a particular choice on how we prescribe the $\mbox{Re}\,\rho$ dependence, the smearing is equivalent to an exactly marginal deformation which preserves the Boltzmann form of the partition function as well as locality and unitarity of the world sheet dynamics.  This is highly unexpected. As we stated earlier, smearing over parameter space generally causes locality and unitarity to breakdown. 

One can in fact think of smearing over the target space $T^2$ as the smearing of space-time $T^2$ for a string theory with CFT being part of the world sheet theory. In this way, the smearing of the target space $T^2$ is like $T \bar T$ deforming string theory. 

The organization of this article is as follows. In section \ref{sec2}, we will review the computation of vacuum amplitude of CFT on world sheet $T^2$ and target space $T^2$. In section \ref{sec3}, we will review the integral kernel for $T \bar T$ deformation originally derived in \cite{Dubovsky:2018bmo}. In section \ref{sec4}, we will describe the result of applying the $T \bar T$ integral kernel on target space $T^2$. For a certain choice regarding how we handle the real part of $\rho$ moduli, we find that the smearing is equivalent to an exactly marginal deformation. In section \ref{sec5}, we will describe the interpretation of the deformation of section \ref{sec4} in terms of conformal perturbation theory. In section \ref{sec6}, we will offer our comments and discussions.

\section{CFT in $2d$ with $T^2$ target space}\label{sec2}

In this section, we will review certain basic facts and our conventions regarding the $2d$ CFT whose target space is $T^2$. Readers wishing to gloss over the technical details are invited to skip to the next section.

\begin{figure}
\begin{center}
\begin{tikzpicture}[scale=1.1, transform shape]
   \draw [black,thick,->](0,-.5) -- (0,4);
    \draw [black,thick,->](-.7,0) -- (6,0);
      \draw [black,thick,->](0,0) -- (4,1);
      \draw [black,thick,->](0,0) -- (1,3);
  \draw [black,-](0,0) -- (4,1)--(5,4)--(1,3)--(0,0);
                    \draw (2.7,.4) node {\small{$\vec{L}^1$}};
                     \draw (.3,2) node {\small{$\vec{L}^2$}};
                      \draw (-.4,-.3) node {\small{$(0,0)$}};
                       \draw (4.1,.6) node {\small{$(L_1^1,L^1_2)$}};
                        \draw (1,3.35) node {\small{$(L_1^2,L^2_2)$}};
                         \draw (5.5,4.3) node {\small{$(L^1_1+L_1^2,L^1_2+L^2_2)$}};
                      \draw [black,thin,domain=0:14.] plot ({1.5*cos(\x)}, {1.5*sin(\x)});
                       \draw (1.8,.25) node {\small{$\phi$}};
 \end{tikzpicture}
\caption{The target space torus parametrized by the vectors $\vec{L}^{1,2}$. Similar figure appeared in \cite{Dubovsky:2018bmo}. }
  \label{fig1}
 \end{center}
\end{figure}
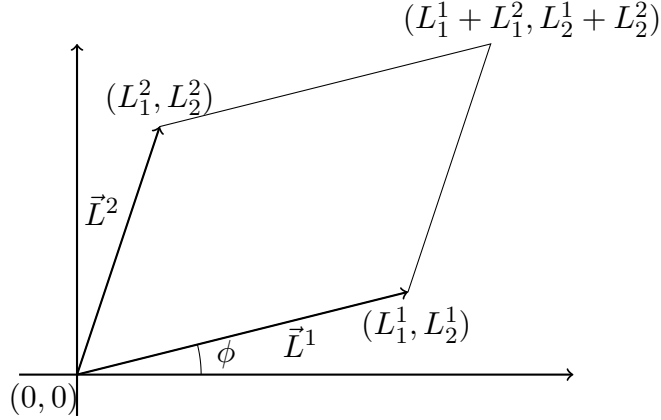

\subsection{Target space $T^2$}

Let us begin by reviewing the geometry of the target space $T^2$. One convenient way to parameterize the target space is by specifying two period vectors
\be \vec L^a = L^a_\mu \hat x^{\mu} ~,\ee
for $a=1,2$. The other label $\mu=1,2$ is the space-time index.  $\hat x^\mu$ are orthogonal unit vectors on the plane spanned by $T^2$. See figure \ref{fig1} for a diagrammatic illustration.   We will treat $L_\mu^a$ as having the dimension of length which is natural when the CFT is viewed as the world sheet dynamics of a string theory. By assuming that there are no preferred direction in the space of $\vec L$, it follows that overall rotation of $\vec L^a$ do not affect the CFT. So the $L_\mu^a$ captures three of the four moduli coordinates. The remaining moduli is the coefficient of the topological term corresponding to the $B$-field in the string theory interpretation as we will elaborate below. 

It will be convenient, at times, to relate the $L_\mu^a$ parameters to the complex and K\"ahler structure parameters. To that end, let us define
\be   \xi = {L_1^2 + i L_2^2 \over L_1^1 + i  L_2^1}~, \qquad (2 \pi)^2  R^2 = (L_1^1)^2 + (L_2^1)^2~, \qquad \phi = \arg(L_1^1+i L_2^1)~.  \label{xiL}\ee
We are using $\xi$ to parameterize the complex structure. $R^2$ sets the overall size and will be related to the K\"ahler structure. As was mentioned earlier, $\phi$ decouples from the dynamics.

\subsection{CFT Action}

Next, we will write down the action for the CFT. We are mostly following the convention of Polchinski \cite{Polchinski:1998rq}.
Let us being by defining the dimensionless world sheet coordinates
\be z = z_1 + i z_2~, \qquad \bar z = z_1 - i z_2~, \ee
which we take to be identified under periodicity
\be z \sim z +2 \pi  \sim z+ 2 \pi \tau ~, \ee
where $\tau$ is the world sheet modular parameter.

We next introduce the target space coordinates $\Gamma_\mu$ with $\mu=1,2$
\be \Gamma_\mu \sim \Gamma_\mu + L^1_\mu \sim \Gamma_\mu + L^2_\mu~, \ee
and define
\be \Gamma = \Gamma_1 + i \Gamma_2~, \qquad \bar \Gamma = \Gamma_1 - i \Gamma_2~. \ee
We can then write the action as
\be S = S_g + S_b~, \ee
where
\be S_g = {1 \over 4 \pi \alpha'} \int d^2 z \, (\partial \bar \Gamma \bar \partial \Gamma + \partial \Gamma \bar \partial \bar \Gamma) \label{Sg}\ee
and
\be S_b = {i b \over 4 \pi \xi_2 R^2} \int d^2 z \, (\partial \bar \Gamma \bar \partial \Gamma - \partial \Gamma \bar \partial \bar \Gamma)~. \ee
The two terms $S_g$ and $S_b$ come from the metric and the $B$-field contribution to the world sheet action. We have adopted the measure convention as given in (2.1.7) \cite{Polchinski:1998rq}
\be d^2 z \equiv dz \, d \bar z = 2 dz_1 d z_2 ~. \ee
A dimensionful constant $\alpha'$ appears in the action in a form that is familiar for string theory application. They can be used to make $\Gamma$ dimensionless or to be set to standard values like $\alpha'=2$ but it is convenient to keep them around for diagnostic purposes.  Also,
\be {\cal A} = L_1^1 L_2^2 - L_1^2 L_2^1~, \ee
is the oriented area of the target space torus.

The $S_b$ term is topological and evaluates to 
\be S_b = 2 \pi i b N~, \ee
for an integer value $N$ corresponding to the winding number (\ref{Nwmwm}) of the embedding of the world sheet torus into the target space torus.

\subsection{Torus Partition Function}

Now let us review the computation of the thermal partition function which can be evaluated as a path integral
\be Z(L^a_\mu/\sqrt{\alpha'}, b; \tau) = \int [D \Gamma] [D \bar \Gamma] \exp[ - S_g - S_b ] ~ . \label{pathintegral} \ee
This quantity is periodic under $b \rightarrow b +1$.  It is therefore natural to also define 
\be Z_N( L^a_\mu/\sqrt{\alpha'}; \tau) = \int_0^{1} db \, e^{- 2 \pi b N}  Z(\tau; L^a_\mu/\sqrt{\alpha'}, b)  \label{ZN}~.\ee

This path integral can be evaluated following the standard textbook methods. One splits the space of field configurations into the topological sector, the zero mode, and the fluctuation modes. The calculation for the case where the target space is $S^1$ instead of $T^2$ can be found in section 8.2 of \cite{Polchinski:1998rq}. For the case where the target space is $T^2$, the analysis of topological sector is summarized in section 3 of \cite{Hashimoto:2019wct}. The resulting closed form expression for the partition function is
\beq \lefteqn{ Z( L^a_\mu/\sqrt{\alpha'}, b; \tau)  }  \label{Z2} \\
&=& \sum_{ m_1,  m_2,  w_1,  w_2} {\xi_2 R^2 \over  \alpha' \tau_2}\exp\left[-{\pi R^2 \over \alpha' \tau_2} | m_1 - \tau  w_1 + \xi ( m_2 - \tau  w_2)|^2 -{2 \pi \xi_2 R^2 \over \alpha'}  N + 2 \pi i b  N\right]  |\eta(\tau)|^{-4}~, \nonumber \eeq
where
\be  N =  w_1  m_2 -  w_2  m_1~, \label{Nwmwm}\ee
and $m_i$ and $w_i$ are the winding numbers of the two one-cycles of the world sheet $T^2$ into $\vec L^a$ cycles and take on integer values. (See equation (3.5) of \cite{Hashimoto:2019wct} for more details.) The relation between $(\xi, \bar \xi, R)$ and $L^a_\mu$ is as given in \eqref{xiL}. 

This thermal partition function can be written in the Boltzmann form by Poisson resumming the sum over $m_i$ as a sum over $n_i$
\be Z( L^a_\mu/\sqrt{\alpha'}, b; \tau)
= \sum_{ n_1,  n_2,  w_1,  w_2}  \exp[-2 \pi \tau_2 \Delta + 2 \pi i (n_1 w_1 + n_2 w_2) \tau_1] |\eta(\tau)|^{-4}~,\label{boltzmann}\ee
with
\be \Delta =\frac{\alpha' ( (n_2+b w_1)-\xi_1 (n_1 -b w_2))^2}{2 \xi_2^2 R^2} +\frac{ \alpha'^2 (n_1-b w_2)^2+R^4 (\xi_1 w_2+w_1)^2}{2
   \alpha' R^2}+\frac{\xi_2^2 R^2 w_2^2}{2 \alpha'}~. \label{Delta}
\ee
This expression is in agreement with (2.4.18) and  (2.4.58) of \cite{Giveon:1994fu} upon defining
\be \rho = b +  {\xi_2 R^2 \over  \alpha'}i ~. \label{rho} \ee
Periodicity $b \rightarrow b+1$ is also manifest in this form.

\section{Smearing Kernel}\label{sec3}

In this section, we will present a specific smearing of the partition function \eqref{Z2} which is the main observation of this note.

The smearing that we consider is largely based on the kernel which appeared as equation (44) of \cite{Dubovsky:2018bmo}. In the context of \cite{Dubovsky:2018bmo}, the $\vec L^a$ parameterized the geometry of the world sheet on which the CFT was defined. This procedure generated the partition function of a $T \bar T$-deformed field theory which is not necessarily conformal.  The difference here is that we will apply the same smearing on the $\vec L^a$ which is parametrizing the geometry of the target space $T^2$. As such, this smearing is not to be interpreted as the $T \bar T$ deformation of the CFT. 

Let us begin by recalling the kernel which appeared as equation (44) of  \cite{Dubovsky:2018bmo} which can be written as
\be I(L_\mu^a,\dot L_\mu^a,\Lambda)  = {\Lambda^2 \A e^{-\Lambda \A} \over (2 \pi)^2 \dot \A}
e^{\Lambda \epsilon^{\mu \nu} \epsilon_{ab} (L_\mu^a \dot L_\nu^b-{1 \over 2} \dot L_\mu^a \dot L_\nu^b)} ~.\label{Dubovsky} \ee
Here, $\dot {\cal A} = \dot L_1^1 \dot L_2^2 - \dot L_1^2 \dot L_2^1$, and $\Lambda$ parameterizes the extent of smearing. The smearing can then be understood as a distribution which we use to integrate over the moduli parameter  $\dot L_\mu^a$. We have used the dot, instead of the bar which was used in  \cite{Dubovsky:2018bmo}, to indicate the integration variable to avoid confusing with the bar implying complex conjugation.

This kernel can also be written in the following form
\be I(L_\mu^a,\dot L_\mu^a,\Lambda) =  {\Lambda^2 \A  \over (2 \pi)^2 \dot \A}
e^{-{1 \over 2}\Lambda \epsilon^{\mu \nu} \epsilon_{ab} (L_\mu^a - \dot L_\mu^a)(L_\nu^b - \dot L_\nu^b)} ~.\label{kernel2}\ee

One important property of $I$ is that it solves the differential relation
\be \partial_{\Lambda^{-1}}I(L_\mu^a,\dot L_\mu^a,\Lambda) = {\cal A}  G^{ab}_{\mu \nu} {\partial \over \partial L_\mu^a}{\partial \over \partial L_\nu^b}  {\cal A}^{-1} I(L_\mu^a,\dot L_\mu^a,\Lambda)~,\label{diffeq} \ee
with
\be G^{ab}_{\mu \nu} {\partial \over \partial L_\mu^a}{\partial \over \partial L_\nu^b}  = \left(\partial_{L_1^1} \partial_{L_2^2} - \partial_{L_1^2} \partial_{L_2^1}\right) ~,\ee
which has the form of a diffusion equation in the space of $L_\mu^a$ where $\Lambda^{-1}$ is playing a role of time. There is however a caveat in that the metric  $G_{ab}^{\mu \nu}$ is not positive definite. Instead, it is a $4 \times 4$ matrix with signature $(2,2)$.  The fact that the signature of $G_{ab}^{\mu \nu}$ is indefinite implies that generic initial condition specified at fixed $\Lambda^{-1}$ will lead to a singularity (of the type one typically encounters when diffusion process evolves backward in time) at some finite $\Lambda^{-1}$. However, we will find that there are certain special initial conditions for which the flow is singularity free. 

Let us summarize few other properties of kernel \eqref{kernel2}.

\begin{enumerate}
\item In the limit $\Lambda \rightarrow \infty$, The kernel \eqref{kernel2} do not localize to a delta function.
\be I(L_\mu^a, \dot{L}_\mu^a,\Lambda) \ne  \delta^4 (L_\mu^a - \dot{ L}_\mu^a) ~. \ee
This is because the signature of $G_{ab}^{\mu \nu}$ is not positive definite. One way to demonstrate this explicitly is to act with the kernel on $\delta^4( L_\mu^a - \dot L_\mu^a)$ and send $\Lambda^{-1} \rightarrow 0$. 

\item The kernel is invariant under
\be L_\mu^a \rightarrow M^a{}_b L_\mu^b, \qquad   \dot L_\mu^a \rightarrow M^a{}_b  \dot L_\mu^b~, \qquad \Lambda = \mbox {fixed}~, \ee
for $M^a{}_b \in SL(2,Z)$. This is a manifestation of the $SL(2,Z) \times SL(2,Z)$ T-duality of a CFT with target space $T^2$. 

\item The kernel \eqref{Dubovsky} respects a composition rule. 
\be \int {d^4 \dot  L} \, I(L_\mu^a, \dot L_\mu^a,\Lambda) I( \dot L_{\mu}^a, \ddot L_\mu^a,\Lambda') =  I\left(L_\mu^a,\ddot L_\mu^a,{\Lambda \Lambda' \over \Lambda+\Lambda'}\right)~. \ee
\end{enumerate}
In order to arrive at statement 3, one uses the following formula for performing a Gaussian integral
\be \int d^4  L \exp \left[- \pi ( L_\mu^a + X_\mu^a) A_{ab}^{\mu \nu} ( L_\nu^b + X_\nu^b)  + B \right]
= {1 \over \sqrt{\det A}} \exp[B]~. \label{gaussian} \ee
Since the signature of $A_{ab}^{\mu \nu}$ is not positive definite some of the Gaussian integral is  unbounded. The integral \eqref{gaussian} is to be understood as being evaluated via analytic continuation in $A_{ab}^{\mu \nu}$ or in the contour of integration in complex $ L$ space. 

\section{Smearing of target space $T^2$ using the $T \bar T$ kernel}\label{sec4}

Before acting with the kernel \eqref{Dubovsky} on the partition function \eqref{Z2}, it is useful to examine how it acts on {\it part} of \eqref{Z2}. Let
\beq  \lefteqn{\tilde Z_N( L^a_\mu/\sqrt{\alpha'}; \tau)
\equiv \exp\left[  {2 \pi \xi_2 R^2 \over \alpha'}  N \right] |\eta(\tau)|^4 Z_N ( L^a_\mu/\sqrt{\alpha'}; \tau) } \\
& = & {  {\cal A} \over (2 \pi)^2 \alpha' \tau_2}  \exp \left[-{\pi \over  (2 \pi)^2  \alpha' \tau_2}\left|  ( L_1^1 +i  L_2^1)(m_1-\tau w_1)+(
   L_1^2+i L_2^2) (m_2-\tau w_2)) \right|^2 \right]~. \nonumber \eeq
It can then be shown using  \eqref{gaussian} that
\beq \lefteqn{ \int {d^4 \dot L} \  I(L_\mu^a, \dot L_\mu^a,\Lambda) \tilde Z_N( \dot L^a_\mu/\sqrt{\alpha'}; \tau) } \cr
& = &  {{\cal A} \over (2 \pi)^2  (\alpha'- ( \pi  \Lambda)^{-1}N) \tau_2} \cr
&& \quad   \exp \left[-{\pi \over  (\alpha' - ( \pi   \Lambda)^{-1}N)\tau_2}\left|  ( L_1^1 +i  L_2^1)(m_1-\tau w_1)+(
  L_1^2+i L_2^2) (m_2-\tau w_2)) \right|^2 \right] \cr
  & = & \tilde Z_N (L^a_\mu/\sqrt{\alpha' - ( \pi \Lambda)^{-1} N};\tau)~. \label{shiftalpha}  \eeq
  
This is a remarkable result in that the smearing by \eqref{Dubovsky} appears to simply shift $\alpha' \rightarrow \alpha' - ( \pi \Lambda)^{-1} N$ of $\tilde Z_N$. This leads us to introduce a new parameter $\mu$ and define
\be \Lambda_N = -{N \over \pi \mu} \label{LambdaN} ~ ,\ee
as the deformation parameter in sector with fixed value of $N$.

It is also worth noting that the $\Lambda^{-1} \rightarrow 0$ limit of the kernel acting on $\tilde Z_N$ is well defined and in fact acts like an identity operation. This is a consequence of the fact that $\tilde Z_N$ is suitably delocalized for the diffusion with positive and negative diffusion coefficient to be well defined globally. This is the singularity free initial condition to which we alluded to earlier in this section.

At this point, we have assembled all the ingredients necessary to define the smearing that we are after. Our proposal is to define a smearing factor
\beq \lefteqn{ F(L,b;  \dot L, \dot b; \mu; \alpha' )} \label{final1} \\
& \equiv& \sum_{N=-\infty}^\infty e^{ \pi i  b N} \exp\left[-{\pi \xi_2  R^2 \over (\alpha' +\mu ) } N \right] I(L_\mu^a,\dot L_\mu^a, \Lambda = \Lambda_N = -{ N \over  \pi \mu})  \exp\left[{2 \pi \dot \xi_2 \dot R^2 \over \alpha'} N \right] e^{-2 \pi i \dot b N}~, \nonumber \eeq
to smear the partition function as follows
\be Z(L^a_\mu/\sqrt{\alpha' +  \mu},b; \tau) = \int_0^1 d\dot b  \int d^4 \dot L\,  F(L,b;  \dot L, \dot b; \mu; \alpha') \, Z(\dot L^a_\mu/\sqrt{\alpha' },\dot b; \tau)~. \label{final2} \ee

Equations \eqref{final1} and \eqref{final2} is the main result of this note. It consists of constructing a weighing  factor  $F$ over the moduli space parameters $\dot L^a_\mu/\sqrt{\alpha'}$ and $\dot b$ which depends on $L^a_\mu/\sqrt{\alpha'}$, $b$, $\mu/\alpha'$ which when used to  smear the partition function depending on moduli parameters $\dot L^a_\mu/\sqrt{\alpha'}$ and $\dot b$, gives rise to an expression which is interpretable as the same partition function that we started with but with the moduli fixed at $L^a_\mu / \sqrt{\alpha' + \mu}$ and $b$. 

There is one important disclaimer that warrants explanation in  expressions \eqref{final1} and \eqref{final2}. The operation consists of expanding the partition function into contributions with fixed $N$, and then deforming by $\Lambda = \Lambda_N = - {N/  \pi \mu}$. However, this factor of $N$ is accompanied by a factor of compensating $1/N$. This means that for $N=0$, special consideration is necessary. The interpretation that $\alpha'$ is shifted by $\mu$ as is indicated in \eqref{final2} is predicated on setting 
\be {N  \over N}=1~, \label{NoverNis1} \ee 
where they appear. 

The most salient feature of \eqref{final1} and \eqref{final2} is that the smeared partition function is interpretable as a partition function of a conventional, unitary, conformal field theory.  Regardless of the lack of exoticness in the form of the smeared partition function, this example suffices to establish a point that some smearing preserves unitarity. 

\section{Smearing as an exactly marginal deformation}\label{sec5}

In section \ref{sec3}, we inferred the conclusion that the smearing \eqref{final1} leads to the rescaling of $L_\mu^a/\sqrt{\alpha'}$ from the observation \eqref{shiftalpha} which is a property of the closed form expression for the partition function \eqref{Z2}.  We will now show that this conclusion could have been inferred at the level of the path integral \eqref{pathintegral} and the differential relation \eqref{diffeq}.

To see this, let us begin by scaling the field variables
\be \Gamma = 2 \pi  R(\gamma_1 + \xi \gamma_2)~, \qquad \bar \Gamma = 2 \pi  R (\gamma_1 + \bar \xi \gamma_2) ~ , \ee
so that the periodicity of $\gamma_{1,2}$ variables are
\be \gamma_{1,2} \sim \gamma_{1,2} + 1 = \gamma_{1,2} + i~, \ee
is not dependent on $L^a_\mu$. The path integral expression then becomes
\be Z(L_\mu^a,b) = \int [D \gamma] [D \bar \gamma] \exp \left[ -S_g - S_b \right]~, \label{funcpart} \ee
where
\begin{eqnarray}\label{Sundef}
S_g &=&\frac{1}{2\pi\alpha'} \int d^2z \Big{[}\left((L^1_1)^2+(L^1_2)^2\right)\partial \gamma_1\bar{\partial}\gamma_1+(L^1_1L^2_1+L^1_2L^2_2)\partial\gamma_1\bar{\partial}\gamma_2\\ \nonumber
&&+(L^1_1L_2^1+L_1^2L^2_2)\partial\gamma_2\bar{\partial}\gamma_1+\left((L^2_1)^2+(L^2_2)^2\right)\partial\gamma_2\bar{\partial}\gamma_2\Big{]}~,
\end{eqnarray}
and
\begin{eqnarray}
S_b=2\pi b\int d^2z(\partial\gamma_2\bar{\partial}\gamma_1-\partial\gamma_1\bar{\partial}\gamma_2)~,
\end{eqnarray}
where all the dependence on $L_\mu^a$ is contained explicitly in the action.

Now, using  \ref{final1} one can show that the effect of infinitesimal smearing on $Z_N$ defined in \eqref{ZN} is 
\be \delta Z_N =  \left({2 \pi N \xi^2 R^2 \over \alpha'^2}  \delta \mu+ e^{-{2 \pi N \xi_2 R^2 \over \alpha'}} {\cal A}   {\pi  \delta \mu \over N}  \left(\partial_{ L_1^1} \partial_{ L_2^2} - \partial_{ L_1^2} \partial_{ L_2^1}\right)  {\cal A}^{-1} e^{2 \pi N \xi^2 R^2 \over \alpha'}\right) Z_N~. \label{smalldef} \ee

Acting on the path integral expression \eqref{funcpart}, we find that
\beq \delta Z  &=& \int [D \Gamma] [D \bar \Gamma] \sum_N e^{-S_N} { \delta \mu \over N} \left(\rule{0ex}{3ex} {i  \over    \alpha'  } (G_{12} - G_{21})
S(\gamma, \bar \gamma) \right. \cr
&& +{1 \over  \pi \alpha'} (G_{12} G_{21} - G_{11} G_{22} )( L_2^1 L_1^2 - L_1^1 L_2^2)  +  (\mbox{terms linear in $G_{ij}$}) \cr
&& \left. + {2 \pi N^2 \xi_2 R^2 \over \alpha'^2} + {(G_{12}-G_{21})^2  \over 2 \pi \alpha'^2} ( L_2^1 L_1^2 - L_1^1 L_2^2) \rule{0ex}{3ex} \right)~,  \label{deltaZ}
\eeq
where we introduced a notation
\be G_{ab}=\int d^2 z \, \partial \gamma_a \bar \partial \gamma_b~, \ee 
and
\be e^{-S_N} = \int db\,  e^{i b N -S(b)} ~. \ee
There are terms linear and quadratic in $G_{ij}$ depending on how the two derivatives with respect to $L$ act. 
The linear term is somewhat complicated and is given by 
\beq \lefteqn{(\mbox{terms linear in $G_{ij}$})} \label{linear} \\
& =&{G_{11}\left((L_1^1)^2+(L_2^1)^2\right)
+G_{12}(L_1^1 L_1^2+L_2^1 L_2^1)+G_{21} (L_1^1 L_1^2+L_2^1 L_2^2)+G_{22}
\left((L_1^2)^2+(L_2^2)^2\right) \over  \pi \alpha' (L_1^1 L_2^2 - L_2^1 L_1^2)}~.  \nonumber \eeq
The right hand side of \eqref{deltaZ} can be interpreted as a path integral computation of expectation value of products of $G_{ab}$'s. Although highly tedious, it can be shown that the second and the third lines on the right hand side of \eqref{deltaZ} individually vanishes. To facilitate the analysis, it is useful to first consider the case where $L_1^2  = L_2^1 = 0$. In the course of this computation, one encounters contact singularities where products of  $\gamma_\mu(z, \bar z)$ are on coincident points. We regulated these singularities by evaluating the fluctuating part of the functional integral in momentum modes. The contact singularities are manifested as the divergence in sum over momentum modes, which we rendered finite using the zeta function regularization procedure. For the sake of completeness, details of the regularization procedure (for the case of single scalar to keep the discussion compact) is contained in  appendix \ref{appA}.

So, the only non-vanishing contribution from the right hand side of \eqref{deltaZ} is the first line, which using
\be N =  i  (G_{12} -G_{21})~ ,\ee 
simply becomes\footnote{Note that $N/N=1$ must be applied here including when $N=0$ as was done in \eqref{NoverNis1}.}
\be \delta Z  = \int [D \Gamma] [D \bar \Gamma] \sum_N e^{-S_N} \delta \mu
S(\gamma, \bar \gamma)  ~. \label{marginal} \ee
This expression is interpretable as the statement that smearing \eqref{final1} infinitesimally is equivalent to an exactly marginal deformation shifting the size of the target space torus infinitesimally. This deformation is integrable in that it applies for taking torus of arbitrary size as the starting point. Since all that we are doing is a formal manipulation of the smearing \eqref{final1} acting on the partition function \eqref{Z2} expressed in the path integral from \eqref{funcpart}, this conclusion was inevitable and that the analysis presented in this subsection can be viewed as a simple (albeit rather tedious) diagnostic of the conclusion \eqref{final2}. 

It is useful, nonetheless, to relate the smearing \eqref{final1} with exactly marginal deformation as is done in \eqref{marginal} in thinking about observables other than the partition function.  It is quite natural to expect that the natural observables of CFT smeared according to \eqref{final1} is the correlation function of  operators of definite dimension of the marginally deformed CFT. It seems highly unlikely, however, that these observables are accessible by simply smearing the generating function of the form
\be Z(L_\mu^a, b;  J_i(z, \bar z); \tau) = \langle e^{i \int d^2 z\,  {\cal O}_i (z,\bar z) J_i(z,\bar z) } \rangle~, \ee
unless the smearing also acts on the source function $J_i(z,\bar z)$ so that they continue to couple to the operators of definite dimension after the smearing. Otherwise, the $J_i$ will likely couple to some linear combination of operators which one must then be disentangled. Carrying out the analysis of this type appears to be quite cumbersome. 

\section{Comments and Discussions}\label{sec6}

This article described a concrete weight function \eqref{final1} for smearing a CFT$_2$ with target space $T^2$ over its Narain moduli space.  For fixed $L^a_\mu$, $b$, and $\mu$, \eqref{final1} defines a distribution over $\dot L_\mu^a$ and $\dot b$ which is different than the uniform measure considered in \cite{Maloney:2020nni,Afkhami-Jeddi:2020ezh,Datta:2021ftn,Benjamin:2021wzr}. The weight function \eqref{final1} was obtained by  applying the integral kernel which usually acts on the world sheet moduli for a torus partition function when performing a $T \bar T$ deformation, and acting instead on the moduli parameters of the target space torus, with some additional detail on how to handle the target space $B$ field. The result of this smearing is \eqref{final2} which turned out to be interpretable as rescaling of the volume $L_\mu^a \rightarrow \sqrt{1 + \mu /\alpha'} L_\mu^a$. This was somewhat unexpected. Generically, smearing over a parameter of a field theory leads to loss of Boltzmann expansion of the partition function. The smearing constructed in \eqref{final1} and \eqref{final2} is a rare exception to this general expectation. 

Smearing field theories generally leads to a loss of locality. A good example is $T \bar T$ deformation itself. CFTs defined on world sheet $T^2$ are physical probes of $T^2$ where the algebra of fields and differential operators encodes the geometric data of $T^2$ in the sense of  \cite{Witten:1985cc,Connes:1994yd}.  A different way to probe a geometry physically is to treat it as the target space. Smearing over the target space is expected to introduce some non-locality. However, when a geometry is probed as a target space of a CFT in two dimensions, the CFT only probes the geometry non-locally. One quick way to see this is to recall that these CFTs respect T-duality. Another way to think about it is that the CFT is a stringy probe of target space geometry and as such has an intrinsic non-locality scale associated with the string tension. In this sense, it is self consistent that  smearing (\ref{final1}) is equivalent to a marginal deformation. The  probe is non-local in target space before and after the smearing.

In this article, we focused mostly on the properties of smearing (\ref{final1}). It is not too difficult to consider slight modifications. For instance, instead of \eqref{LambdaN}, we could simply set 
\be \Lambda_N = -{1 \over  \pi \mu} \ . 
\ee
This will give rise to a partition function where one assigns a different moduli for each sector labeled by $N$. This will cause the partition function not take on Boltzmann form \eqref{boltzmann}. Another possibility is to smear only over the complex or the K\"ahler moduli of the target space using the kernel of \cite{Hashimoto:2019wct}. In fact, as was shown in \cite{Dijkgraaf:1987jt,Wendland:2000ye,Benjamin:2021ygh}, $c=2$ Narayan CFT with target space $T^2$ exhibits a triality in the dependence on world sheet moduli $\tau$, target space complex structure moduli $\xi$, and the target space K\"ahler structure moduli $\rho$.\footnote{See specifically  Theorem 4.3.1 in \cite{Wendland:2000ye} or the equation (3.31) of
\cite{Benjamin:2021ygh}. We thank the referee for bringing these facts to our attention.} While we considered the smearing (\ref{final1}) as acting on $\xi$ and $\rho$, this triality implies that it can act on any pair out of the set consisting of $\{\tau, \xi, \rho\}$ and still lead to an ordinary partition function of a CFT. This appears to imply that $T \bar T$ deformation combined with a smearing of $\tau$ or $\rho$ can keep the theory a local one. It would be very interesting to explore other variations along these lines.

It is straight forward to generalize \eqref{final1} to the case where the target space is a product  $T^{2n} = T^2 \times T^2\times  \ldots \times T^2$. One simply acts with \eqref{final1} one $T^2$ at a time. Such an approach clearly do not treat the $2n$ cycles of $T^{2n}$ on equal footing. Because of the way in which the $B$ field plays a critical role in \eqref{final1}, it is not easy to imagine a natural generation that do not implcitly reference two cycles. This may be related to the fact that $T \bar T$ deformation continues to resist elegant generalizations to dimensions other than two. Another interesting question  to explore is whether other marginal deformations could be realized as a smearing.  The smearing \eqref{final1} appears to provide a concrete example which one can use as a foundation for future investigations on related topics.

\section*{Acknowledgements}
We would like to thank D. Kutasov for collaboration at early stages of this work and helpful discussions. 
The work of SC is supported by the Infosys Endowment for the study of the Quantum Structure of Spacetime.

\appendix

\section{Path Integral and  the $\zeta$-function Regularization \label{appA}}

In this appendix, we will review the computation of quantities such as 
\be \left\langle \int d^2 z \, \partial y (z, \bar z) \bar \partial y (z, \bar z) \right\rangle, \qquad 
\left\langle \int d^2 z \int d^2 w  \, \partial y (z, \bar z) \bar \partial y (w, \bar w)  \partial y (w, \bar w) \bar \partial y (z, \bar z)\right\rangle~ .
 \ee
The issue here is the handling of contribution to these expectation values when $y$ are coincident when integrating over $z$ and $w$. Such a contribution is strictly speaking singular. We will describe how a finite contribution is extracted for these quantities using the $\zeta$ function regularization method.

To keep the discussion simple, we will work with single compact scalar with action
\be S = {1 \over 2 \pi \alpha'} \int d^2 z\,  \partial y \bar \partial y~, \ee
where we are following the notational convention of section 2.1 of \cite{Polchinski:1998rq}. Scalar field $y$ is identified periodically under
\be y \sim y + 2 \pi R~. \ee
The periodicity of the world sheet coordinate is
\be z \sim z + 2\pi  \sim z + 2 \pi \tau~ . \ee
Let us also take the world sheet torus to be right angled so that $\tau_1=0$ to keep the discussion simple.  Then, $y$ can be expanded into 
\be y =  w R z_1 +  {m R z_2 \over \tau_2} + y_0 + \sum_{(m_1,m_2) \ne (0,0)} c(m_1,m_2) e^{i m_1  z_1 + i m_2 z_2 / \tau_2}~. \ee
The variables $(m,w)$ labels the topological sector. $y_0$ is the zero mode, and $c(m_1,m_2)$ is the coefficient of the fluctuating component. 

Then, 
\beq\int d^2 z \, \partial y \bar \partial y &=& \left( 2 \pi ^2 R^2 \tau_2
   w^2-\frac{2 \pi ^2 m^2
   R^2}{\tau_2} \right) \cr
   &&  + \sum_{(m_1, m_2) \ne (0,0)} 2 \pi^2 \left(\tau_2 m_1^2 + {m_2^2 \over \tau_2}\right) c(m_1,m_2) c(-m_1,-m_2) ~.
  \label{modeexpand} \eeq
The first term  of \eqref{modeexpand} is the ``topological term'' 
\be (\mbox{topological}) = \left( 2 \pi ^2 R^2 \tau_2
   w^2-\frac{2 \pi ^2 m^2
   R^2}{\tau_2} \right)~, \ee
whereas the second term is the contribution of the fluctuating modes
\be (\mbox{fluctuating}) = \sum_{(m_1, m_2) \ne (0,0)} 2 \pi^2 \left(\tau_2 m_1^2 + {m_2^2 \over \tau_2}\right) c(m_1,m_2) c(-m_1,-m_2)~,  \ee

Our task is to compute the expectation value for an expression of this type weighted by the action. 
The topological sector will give rise to to the usual sum over momentum and winding via Poisson resummation.  The contribution from the fluctuating modes will give
\be (\mbox{fluctuating}) = \sum_{(m_1, m_2) \ne (0,0)} {1 \over 2} (2 \pi \alpha') = -{1 \over 2}(2 \pi \alpha')~.  \ee
To do this sum, it is convenient to sum over contribution from a) sum over $m_1 \ge 1$ and $m_2=0$ and b) summing first over $m_1 \ge 1$ with some fixed $m_2 \ge 1$, and then summing over $m_2 \ge 1$. All of these sums are formally divergent but can be evaluated using the usual  $\zeta$ function regularized expression 
\be \sum_n {1 \over n^a} = \zeta(a)~ , \qquad \zeta(0) = -{1 \over 2} ~.\ee

All expectation values of the type we encounter can be evaluated along these lines. Using these ingredients, one can show the cancellation, line by line, for the second and the third lines of \eqref{deltaZ}.

\providecommand{\href}[2]{#2}\begingroup\raggedright\endgroup

\end{document}